\documentclass{jfm}

\usepackage{graphicx}
\usepackage{epstopdf}
\usepackage{natbib}
\usepackage{xcolor}

\ifCUPmtlplainloaded \else
  \checkfont{eurm10}
  \iffontfound
    \IfFileExists{upmath.sty}
      {\typeout{^^JFound AMS Euler Roman fonts on the system,
                   using the 'upmath' package.^^J}%
       \usepackage{upmath}}
      {\typeout{^^JFound AMS Euler Roman fonts on the system, but you
                   dont seem to have the}%
       \typeout{'upmath' package installed. JFM.cls can take advantage
                 of these fonts,^^Jif you use 'upmath' package.^^J}%
      }
  \else
  \fi
\fi

\ifCUPmtlplainloaded \else
  \checkfont{msam10}
  \iffontfound
    \IfFileExists{amssymb.sty}
      {\typeout{^^JFound AMS Symbol fonts on the system, using the
                'amssymb' package.^^J}%
       \usepackage{amssymb}%

      }{}
  \fi
\fi

\ifCUPmtlplainloaded \else
  \IfFileExists{amsbsy.sty}
    {\typeout{^^JFound the 'amsbsy' package on the system, using it.^^J}%
     \usepackage{amsbsy}}
    {\providecommand\boldsymbol[1]{\mbox{\boldmath $##1$}}}
\fi

\newsavebox{\astrutbox}
\sbox{\astrutbox}{\rule[-5pt]{0pt}{20pt}}

\title[Expanding liquid sheet in  air]{Free radially expanding liquid sheet in  air: time- and space-resolved measurement of  the thickness field}

\author[Clara Vernay, Laurence  Ramos and Christian Ligoure]%
{C. Vernay, %
L. Ramos$\dagger$
and C. Ligoure%
 \thanks{Email addresses for correspondence: christian.ligoure@univ-montp2.fr, laurence.ramos@univ-montp2.fr},\ns}
\affiliation{
Laboratoire Charles Coulomb (L2C), UMR 5221 CNRS/Universit\'e Montpellier 2, F-34095 Montpellier Cedex 05 , France\\[\affilskip]}

\pubyear{2010}
\volume{650}
\pagerange{119--126}
\date{?; revised ?; accepted ?. - To be entered by editorial office}

\begin{document}

\maketitle

\begin{abstract}
The collision of a liquid drop against a small target results in the formation of a thin   liquid sheet that extends  radially until it reaches a maximum diameter. The subsequent retraction is due to the air-liquid surface tension.
We have used a time- and space-resolved technique to measure the thickness field of this class of liquid sheet, based on the grey level measurement of the image of a dyed liquid sheet recorded using a fast camera.  This method enables a precise measurement of the thickness in the range  $(10-450) \, \mathrm{\mu m}$, with a temporal resolution equals to that of the camera. We have measured the evolution with time since impact, $t$, and radial position, $r$, of the thickness, $h(r,t)$, for various drop volumes and impact velocities. Two asymptotic regimes for the expansion of the sheet are evidenced. The scalings of the thickness with $t$ and $r$ measured in the two regimes are those that were predicted in \citet{Rozhkov2004}  fort the short-time regime and \citet{Villermaux2011}  for the long time regime, but never experimentally measured before. Interestingly, our experimental data also evidence the existence of a maximum of the film thickness $h_{\rm{max}}(r)$ at a radial position $r_{\rm{h_{max}}}(t)$ corresponding to the crossover of these two asymptotic regimes. The maximum moves  with a constant velocity of the order of the drop impact velocity, as expected theoretically. Thanks to our visualization technique, we also evidence an azimuthal thickness modulation of the liquid sheets.
\end{abstract}


\section{Introduction }
\label{intro}

The abundance of natural, technical  or even societal  phenomena \citep{Attinger2013} accompanying  drop impacts  on solid or liquid surfaces has led to a huge scientific literature since the beginning of the 19th century \citep{Worthington1908}. In spite of their commonness,  the topics of drop impact continue to increasingly attract analytical, numerical and experimental studies, because related physical phenomena are far from being fully understood. The reader can refer to some comprehensive reviews \citep{Rein1993,Yarin2006} or to a more specific recent one by \citet{Marengo2011}.

Only recently has the impact of droplets onto small targets been investigated. A small target is defined as one for which the size is on the same order of magnitude as the size of the impacting drops. \citet{Rozhkov2002,Rozhkov2004} and \citet{Villermaux2011} reported on experimental and theoretical investigations on the impact of a water drop onto a small flat cylinder. These experiments were subsequently extended to complex fluids, i.e., dilute  high molecular weight polymer solutions \citep{Rozhkov2006} or surfactant solutions \citep{Rozhkov2010}. \citet{Juarez2012} presented the effect of the flat target cross-sectional geometry on the splashing symmetry and  instabilities. \citet{Bakshi2007} conducted investigations on the impact of a droplet onto a spherical target whereas \citet{Hung1999} presented results on the impact of water droplets on cylindrical wires of comparable diameters. It was argued \citep{Rozhkov2002} that investigating the collision of a drop against a target of dimension similar to that of the drop can be seen as a model experimental configuration for the impact of a drop on a rough surface when the linear scale of the surface roughness and the size of the drop have the same order of magnitude. This condition resembles the impingement of an ink drop from an ink-jet printer onto paper as the length scale of the droplet is comparable to the roughness size of the paper.

 On a more fundamental point of view, the drop collision on a small target can also be considered as an experimental simplification of the process of drop collision on a solid surface: indeed,  in the small target configuration, the  liquid sheet expands freely in air, avoiding strong interactions with solid wall and significant viscous dissipation. Upon impact, a drop flattens into a radially expanding sheet which impales on the impactor while remaining attached to it. The sheet consists of a free liquid film, which expands and then recedes, and is surrounded by a thicker rim. The maximum diameter of the sheet, $d_{\rm{max}}$, depends on $u_{0}$, the impact velocity, and on $ d_{0}$, the falling drop diameter. Simultaneously to the sheet expansion and retraction, the rim corrugates forming radial ligaments which subsequently destabilize into droplets. The  analysis of the drop fragmentation process has been reported by \citet{Villermaux2011}. Note also that this configuration is the non stationary version of  a  Savart  sheet formed through the impact of a continuous liquid jet on a small disk at high Reynolds number \citep{Clanet2002,Clanet2007,Bremond2007}. Such a free radially liquid sheet expanding in  air and generating fluid threads and droplets at a high speed is also observed when a small volume of fluid is quickly compressed  between two objects \citep{Gart2013}.

\citet{Rozhkov2002,Rozhkov2004} on the one hand and \citet{Villermaux2011} on the other hand proposed a theoretical model to describe the hydrodynamics of the free  radially expanding liquid sheet, that lead to a same radial velocity field, $u(r,t)$ but to different predictions for the radial sheet thickness field, $h(r,t)$ and  the sheet diameter, $d(t)$. The two models intend to describe two different limits of the phenomenon: the first one addresses time scales short compared to the collision time ($\tau_{\rm{coll}}= d_{0}/ u_{0}$) and the second one is explicitly concerned with larger time scales. Both theoretical approaches are based on the resolution of Euler equations in the slender-slope approximation with axisymmetry associated with  the resolution of linear momentum balance  equation for the variable-mass  sheet rim. \citet{Rozhkov2004} considered that the drop impact is equivalent to a cylindrical point source that will feed the expanding sheet at short times. This leads to a velocity field $u(r,t)$ that decreases with $t$ and increases with $r$, whereas the thickness field, $h(r,t)$  increases with $t$ and decreases with $r$. The asymptotic behavior, valid for short times $t<3\tau_{\rm{coll}}$, with $\tau_{\rm{coll}}$, the typical collision time $\tau_{\rm{coll}}= d_{0}/ u_{0}$,  gives $u(r,t)\propto r/t$ and $h(r,t)\propto t/r^3$. By contrast, \citet{Villermaux2011} considered the implicit opposite limit  $t>>\tau_{\rm{coll}}$. They proposed an elegant  solution for $u(r,t)$ and $h(r,t)$ that is a time-dependent adaptation of a steady-state axisymmetric solution of Euler equations for a continuous jet impacting a solid target. This provides a theoretical prediction for the motion within the sheet, $u(r,t)\propto r/t$, similar to that of  \citet{Rozhkov2004}, but a sheet thickness field, $h(r,t)\propto1/rt$, which  differs strongly from Rhozkov predictions.

Up to now, the confrontations of the theoretical models developed by \citet{Rozhkov2004} and \citet{Villermaux2011} were restricted to the indirect comparison of the measurement of the sheet diameter $d(t)$. Indeed, experimental techniques to access time- and space-resolved liquid film thickness \citep{Tibirica2010} had not been yet adapted to the measurement of liquid sheet thickness. In both experimental reports \citep{Rozhkov2004, Villermaux2011}, the sheet diameter time evolution exhibits a maximum in semi-quantitative agreement with theoretical models. However this indirect comparison is clearly not sufficient to discriminate between the different theoretical models. To reach this goal, direct measurements of the hydrodynamic fields are clearly desired. In this optics, \citet{Bakshi2007} measured the profile of a drop impacting onto a small sphere from the direct capture of sequences of side view images of the impact. However, the spatial resolution was too low to quantitatively compare experimental results with theoretical predictions except close to top of the sphere where the drop has impacted its target. To overcome those limitations we have used a robust and sensitive experimental method to measure the thickness field $h(r,t)$ of a  liquid sheet. Thanks to our measurements, a comparison of the experimental evolution of $h(r,t)$  for various drop volumes and impact velocities with the theoretical models developed by \citet{Rozhkov2004} and \citet{Villermaux2011} is possible.

The paper is organized as follows. In section  \ref{Materials}, we present the method we have developed to measure the thickness field, $h(r,t)$. The experimental results are reported and compared to the theoretical predictions in section \ref{Results}. Finally, we will briefly conclude and provide perspectives to our work in section \ref{Conclusion and perspectives}.

\section{Setup and experimental methods}
\label{Materials}

\subsection{Drop impact experimental setup}

\begin{figure}
\centerline{\includegraphics[height=6cm,width=7cm]{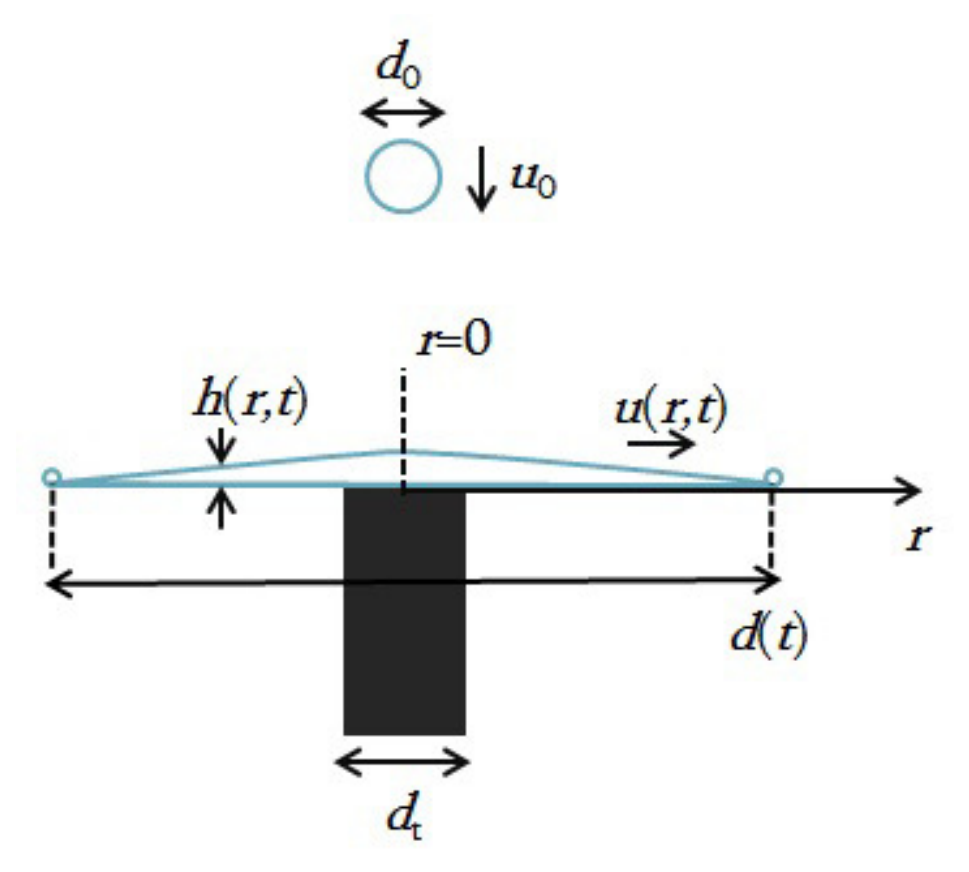}}
  \caption{Sketch of the drop impact experimental setup. The experimental parameters are defined on the sketch.}
\label{fig:fig1}
\end{figure}

Our experimental setup is adapted from the original setup designed by \citet{Rozhkov2002}. A scheme of the set-up is shown in figure~\ref{fig:fig1}. In brief, we let a liquid drop fall on a solid cylindrical target of diameter slightly larger than that of the drop. The liquid drop is injected with a syringe pump and a needle positioned vertically with respect to the target. The impact velocity of the drop, $u_0$, is set by the height of fall of the drop and the drop diameter, $d_0$, is set by the needle diameter. Three different drop diameters ($d_0=3.0, 3.7$ and $4.8$ mm) and two impact velocities ($u_0=2.8$ and $4.0$ m/s) are studied. In all experiments the ratio between the drop diameter and the target diameter is kept constant at $0.6$. The Weber number, $We=\frac{\rho u_0^2 d_0}{\gamma}$, respectively the Reynolds number, $Re=\frac{\rho u_0 d_0}{\eta}$, where $\gamma$ is the fluid surface tension, $\rho$ its density and $\eta$ its viscosity, varies from $320$ to $810$, respectively from $8400$ to $19000$. The target is an aluminium cylinder coated on the top with a glass lamella. The glass is rendered hydrophilic thanks to a plasma treatment with a plasma gun (Corona Surface Treater, Electro-Technic products), which suppresses the dewetting on the surface \citep{Marengo2011, Dhiman2009}. A thin aluminium coaxial cylinder is added around the target as suggested by \citet{Villermaux2011} in order to keep the sheet ejection angle equals to 90$^{\circ}$  (with respect to the target axis) and obtain a horizontal and flat sheet. A vertical tube which runs from the needle to slightly above the target is mounted to avoid the effect of draughts and ensure a centered impact on the target. The target is mounted on a transparent plexiglass floor lighted from below by a high luminosity backlight (Phlox High Bright LED Backlights). The drop impact is recorded from the top with a weak angle using a Phantom V7.3 camera operating at $9708$ pictures per second and with a resolution of $576$ x $552$ pixels. Side views are also recorded to check the flatness of the liquid sheet and measure the impact velocity.

\subsection{Sheet thickness measurement}

Measuring liquid film thicknesses on the micrometer scale range is a problem of great interest for various industrial and academic applications. Different thickness measurement techniques have been developed over the last century. \citet{Tibirica2010} presented a comprehensive literature review of the experimental techniques developed to measure liquid film thicknesses ranging from micrometers to centimeters. Available methods can be classified based on the physical principles, acoustic, electrical or optical, involved for the measurements. In particular, numerous optical methods have been developed, including interferometry methods \citep{Ohyama1988}, fluorescence intensity methods \citep{Makarytchev2001}, methods based on imaging a diffusive liquid \citep{Berhanu2013} or light absorption methods \citep{Lilleleht1961a, Kim2005}. \citet{Mouza2000} compared two methods to measure liquid film thickness, a photometric technique based on the absorption of light passing through a layer of dyed liquid and a conductance measurement and found a very good agreement between these two methods. The authors also pointed out the advantages of optical methods based on light absorption over conductance method. Optical methods are non-intrusive, suitable for non-conducting fluids and can display sensitivity to thin films and high spatial resolution.

\begin{figure}
\centerline{\includegraphics[height=12cm,width=17cm]{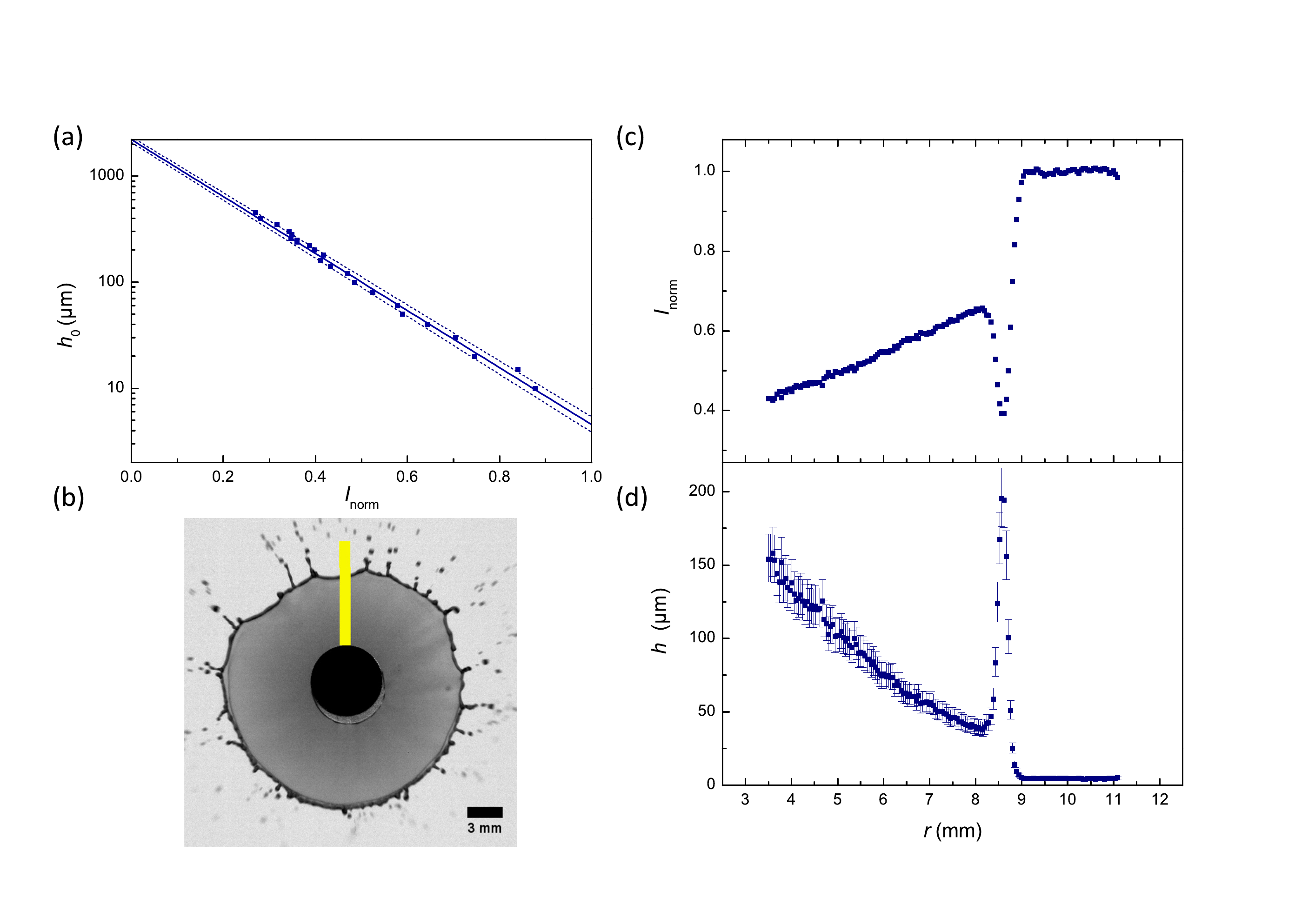}}
  \caption{Calibration process (a) Calibration curve, film thickness, $h_0$, as a function of the normalized intensity. The full line is the best fit of the experimental data with the functional form $h_0=L\exp(-BI)$, where $L=2207 \mu$m and $B=6.184$ are fitting parameters, and the dashed lines are the two envelope curves deduced from the standard deviation of the fit parameters. (b) Image of an expanding dyed liquid sheet, at time $t=2.70$ ms. The drop diameter is $d_0=3.7$ mm and the impact velocity is $u_0=4.0$ m/s; (c) Normalized intensity profile as a function of the radial distance from the target center as measured along the line shown in (b); (d) Thickness profile as deduced from the calibration curve shown in (a) along the line shown in (b). Error bars are deduced from the ones of the calibration curve (a).}
\label{fig:fig2}
\end{figure}

Here, we present a method to determine the thickness of liquid sheets, which is based on the light absorption by a dyed liquid. The method that we have developed relies on the utilization of a dye solution to visualize the variation of the thickness field $h(r,t)$, with $t$ the time elapsed since the impact and $r$ the radial position. We have chosen a dye (erioglaucine disodium salt, purchased from Sigma-Aldrich) which exhibits a very high molar extinction coefficient ($\varepsilon > 80000  \rm{M}^{-1} \rm{cm}^{-1}$ for wavelength  in the range $(627-637)$ nm). It is here diluted in milliQ water at a concentration of $2.5$ g/L. This concentration corresponds to the minimum dye concentration required to obtain a sufficient contrast between the air and the liquid sheet and hence to accurately quantify thickness variations. We have checked that the addition of dye at this concentration has no effect on the hydrodynamic of the water sheet. The principle of the measurement is to correlate the grey level of each pixel of the image of a dyed liquid sheet to the local thickness of the liquid sheet. The key point for the success of this method is a very meticulous calibration performed using liquid films of precise known thicknesses. Liquid films of controlled thickness from $10$ to $450$ $\mu$m are obtained by introducing various controlled volumes of liquid between two microscope coverslips ($22 \times 22 \, \rm{mm}^2$) with a micropipette. The uniform thickness of the liquid films ($h_0$) is determined from the volume of liquid ($V_{\rm{liq}}$) and the microscope coverslip area ($A_{\rm{cov}}$): $h_0=\frac{V_{\rm{liq}}}{A_{\rm{cov}}}$. Images of the liquid between the two coverslips are recorded in experimental configurations comparable to that used for the liquid sheet imaging. The contribution of the empty cell (the two microscope coverslips) is taken into account by dividing the grey level of the image of a dyed liquid film in the cell by the grey level of the image of the empty cell. Hence, one defines a normalized intensity, $I_{\rm{norm}}=I/I_{0}$, where $I$ is the intensity measured for the cell filled with liquid and $I_{0}$ is the intensity measured for an empty cell. $I$ and $I_0$ are intensities spatially averaged over all pixels of the cell made with the two coverslips. By construction $I_{\rm{norm}}$ is equal to $1$ for a film thickness equals to zero.  Moreover $I_{\rm{norm}}=I/I_{0}$ tends to $0$ for large thicknesses. The calibration curve, where the film thickness, $h$, is plotted as a function of the normalized intensity, $I_{\rm{norm}}$, is plotted in fig.~\ref{fig:fig2}(a), and shows an exponential decrease of $h$ with $I_{\rm{norm}}$, over the whole range of thicknesses investigated $[(10-450)\, \mu\rm{m}]$. The calibration curve has been obtained by at least three repetitive measurements for each film thickness. The envelope curves represented in fig.~\ref{fig:fig2}(a) corresponds to the standard deviation of the measurements. The different measurements for a given thickness were obtained by slightly modifying the incident light intensity to ensure that a slight change in the illumination conditions does not notably modify the numerical values for the computed normalized intensity. We note that the calibration curve differs from the exponential decrease of the intensity with the thickness predicted by the Beer-Lambert law presumably because of the non-monochromatic light source \citep{Wentworfh1966}. Sequences of images of the drop impacting the target are recorded in the same conditions (illumination, camera settings) as those used for the calibration.

\subsection{Image analysis}
The software ImageJ is used for the image analysis. Firstly, to determine the thickness, we measure for each pixel of polar coordinates $(r,\theta)$ the normalized intensity $I_{\rm{norm}}(r,\theta)$ as the ratio between the measured intensity for a given image of the liquid sheet $I(r,\theta)$ and the measured intensity for an image taken just before the drop impact. The liquid sheet thickness $h(r,\theta)$ is then evaluated thanks to the calibration curve (fig.~\ref{fig:fig2}(a)). The procedure is illustrated in figure~\ref{fig:fig2}. Figure~\ref{fig:fig2}(b) shows a typical image of a liquid sheet. The normalized intensity profile along the radial line shown in fig.~\ref{fig:fig2}(b) (the origin of the radial distance is taken at the center of the target) is plotted in figure~\ref{fig:fig2}(c). The profile displays a non monotonic evolution, with a continuous increase with $r$, which is followed by a sharp drop related to the thicker rim, before reaching a constant intensity equal to $1$ corresponding to the exterior of the liquid sheet. The related thickness profile has been calculated using the calibration curve (fig.~\ref{fig:fig2}(a)) and is shown in figure~\ref{fig:fig2}(d). Here the error bars are deduced from those of the calibration curve.  The peak in the thickness visible in figure \ref{fig:fig2}(d) corresponds to the thicker rim.
In the following, unless noted, the thickness profiles are averaged over all azimuthal angles, $\theta$, to obtain a thickness that only depends on the radial distance, $r$, and on time, $t$. However, a careful observation of figure \ref{fig:fig2}(b) reveals the presence of small fluctuations of the thickness of the liquid sheet. These azimuthal thickness modulations will be discussed in section \ref{Modulation}. Moreover, we systematically exclude the rim from our data analysis because of the irregularities of the contour of the sheet.

Secondly, to quantify the time evolution of the size of the sheet, we locate the rim by binary thresholding the images using ImageJ. This allows one to determine the sheet contours and measure the area of the sheet ($A$).  An apparent diameter for the sheet, $d$, is simply deduced from the area of the sheet according to $d=\sqrt{\frac{4 A}{ \pi}}$.

Thirdly, in order to determine the ejection volume rate at the target periphery, the time evolution of volume of liquid remaining on the target after the impact of the drop is measured thanks to side view images of the drop.  As illustrated in  figure~\ref{fig:fig7}(a-b), for each image we measure the apparent radius $R(z)$ of the liquid drop as a function of the $z$-axis from $z=0$ corresponding to the target and $z_{\rm{max}}$  corresponding to the top of the drop. Thanks to the axisymmetry of the drop crushing on the target, the liquid volume $V$ remaining on the target at each time $t$ is computed using $V=\int_0^{z_{\rm{max}}} \! \pi R(z)^2 \, \mathrm{d}z$.

Finally, azimuthal profile along a circle centered on the center of the target are derived using a MATLAB program.

\section{Experimental results}
\label{Results}

\subsection{Time- and space-resolved measurements of the sheet thickness}

\begin{figure}
\centerline{\includegraphics[height=11cm,width=9cm]{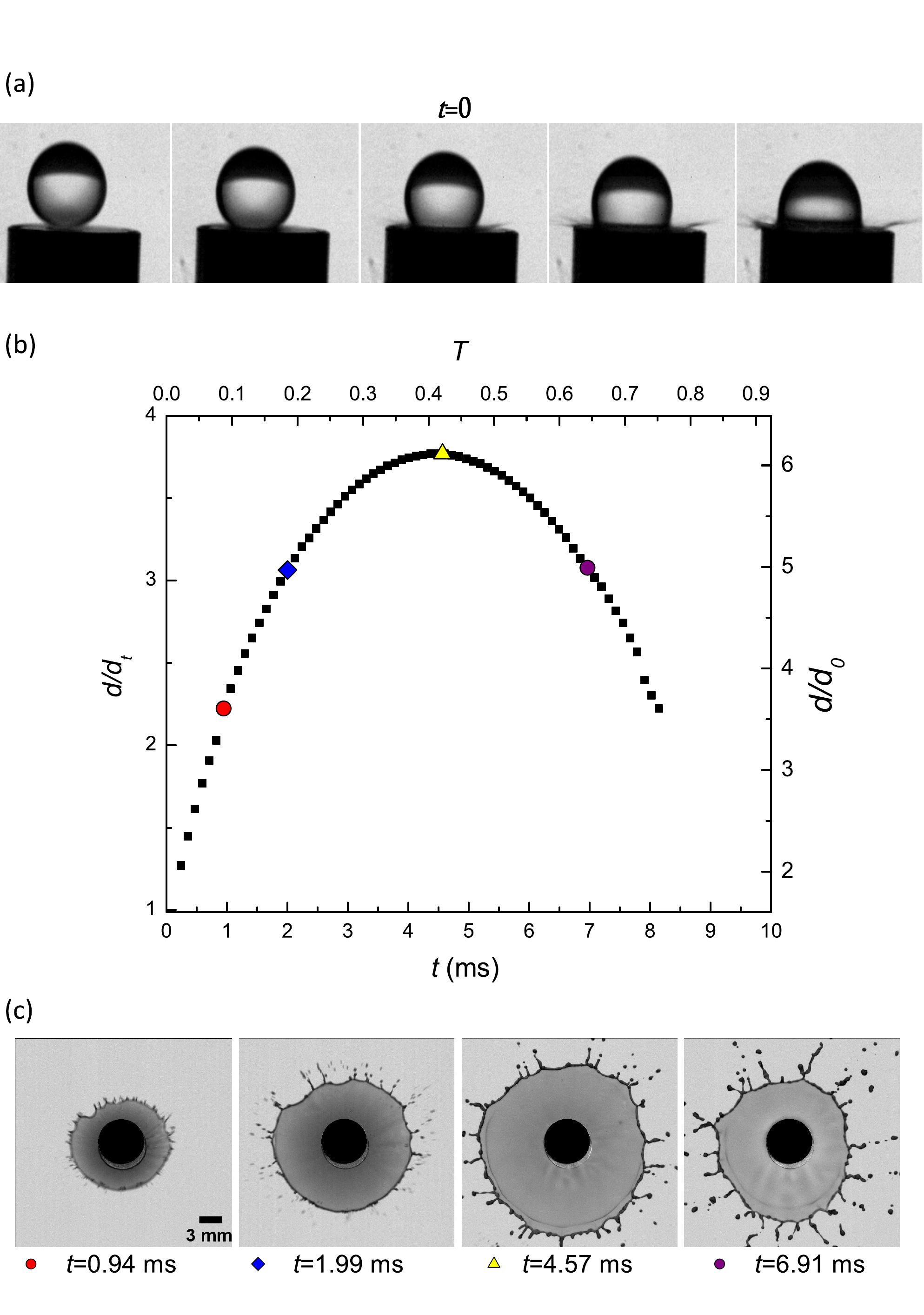}}
  \caption{(a) Sequence of events of the drop just before and after the impact. The time delay between consecutive images is $0.1$ ms. The time $t=0$ corresponds to the impact time of the drop on the target, i.e. the third image in the row. (b) Time evolution of the normalized sheet diameter. The instants corresponding to the four images shown in (c) are highlighted. (c)  Sequence of events of the liquid sheet after drop impact. The drop diameter is $d_0=3.7$ mm, the impact velocity is $u_0=4.0$ m/s, and the target diameter is $6$ mm.}
\label{fig:fig3}
\end{figure}

In a first set of experiments, we fix the size of the drop and the impact velocity. We measure a drop diameter $d_0=3.7$ mm and an impact velocity $u_0=4.0$ m/s, yielding a Weber number $We=810$. Representative pictures of the liquid sheet upon its expansion and retraction are shown in figure~\ref{fig:fig3}. Note that, from the very beginning of the sheet expansion, small droplets are expelled, recalling the microsplash studied by \citep{Thoroddsen2012} upon drop impacts. The time evolution of the normalized sheet diameter, $d/d_0$ (right axis), shows a non-monotonic evolution, reflecting the expansion and retraction regimes. In the following, we take as the origin of time the moment at which the drop impacts the target. In practice, the time $t$=0 is defined as the time when small droplets are expelled from the target, as determined from a top view of the experiment, as shown in figure~\ref{fig:fig3}(a). On the left axis of figure~\ref{fig:fig3}(b) the sheet diameter, $d$, is normalized by the target diameter, $d_t$. We find that at $t=0$ the sheet diameter is closed to the target diameter, ensuring that our determination of the origin of time is reasonable. On the right axis, $d$ is normalized by the drop diameter, $d_0$. In figure~\ref{fig:fig3}(b), we show data both as a function of the natural time, $t$, and as a function of a normalized time $T=t/ \tau$, where $\tau=\sqrt{\frac{\rho d_0^3}{6 \gamma}}$ is proportional to the  capillary time.  Here $\rho=1000 \rm{kg/m}^3$  the density of the dyed water, and $\gamma=71$ mN/m is the air-liquid surface tension. The maximal expansion occurs at a time $t=4.57$ ms ($T=0.42$) and is equal to  $22.62$ mm ($d/d_0=6.11$).  These numbers are consistent with the literature, although the exact shape cannot be fitted satisfyingly with the analytical model developed by \citet{Villermaux2011}. The time at which the sheet reaches its maximum expansion is not the one predicted by the model ($T=1/3$) \citep{Villermaux2011}.This time is neither in agreement with the one ($\frac{4 d_0}{u_0}= 3.7$ ms in our case) found by \citet{Rozhkov2004}. However, it should be highlighted here that we are not  quite in the same experimental conditions than those considered in \citet{Villermaux2011} model ($d_0=d_t$) and in \citet{Rozhkov2002} ($d_0/d_t=1$ and $d_0/d_t=0.72$). This could explain the discrepancy observed in the evolution of the sheet maximum diameter.

\begin{figure}
\centerline{\includegraphics[height=9cm,width=13cm]{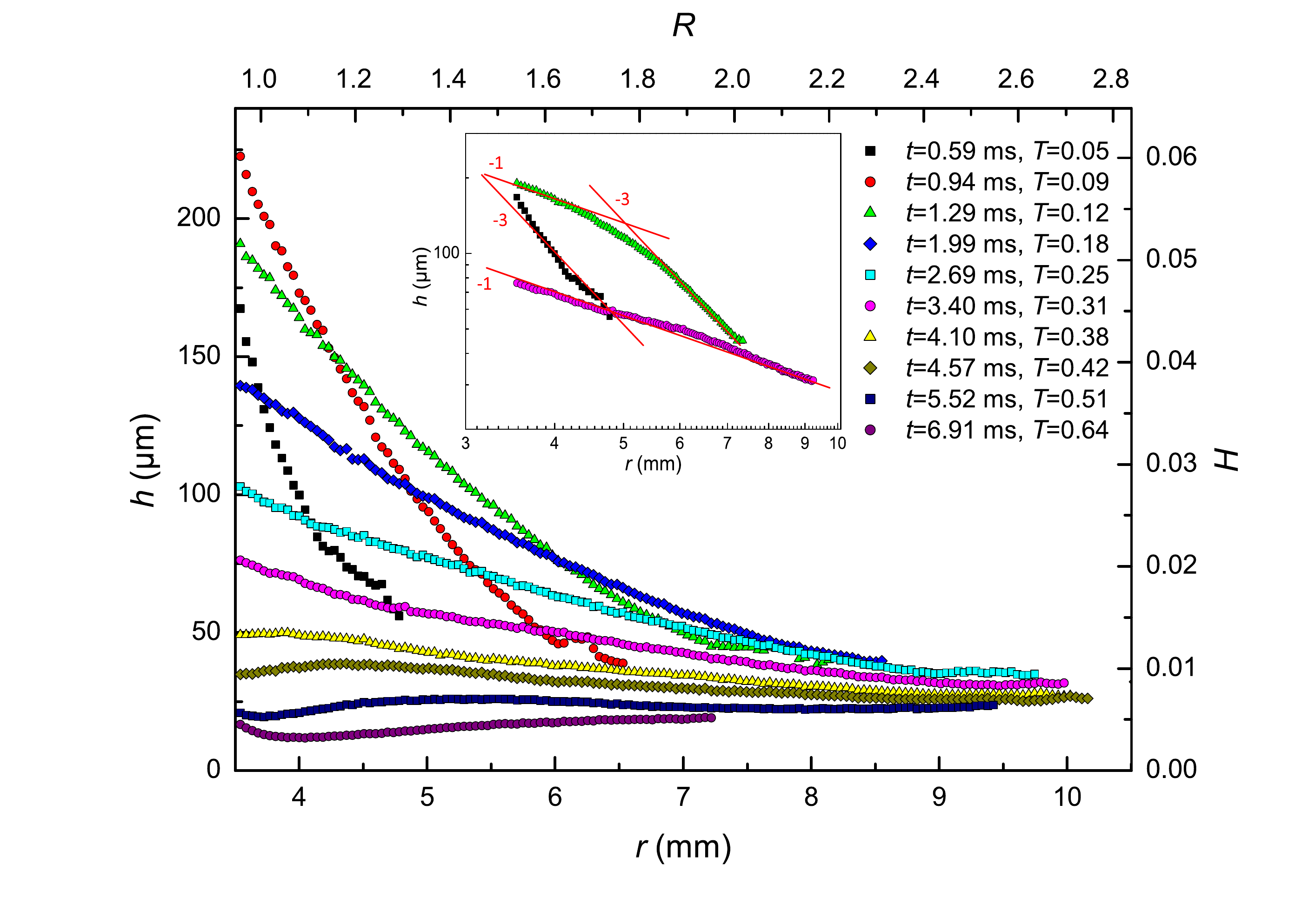}}
  \caption{Thickness of the liquid sheet as a function of radial positions. Actual and normalized data are plotted at different times as indicated in the legend. The drop diameter is $d_0=3.7$ mm and the impact velocity is $u_0=4.0$ m/s. (Inset) Representative data in a log/log scale to highlight the two asymptotic scaling regimes.}
\label{fig:fig4}
\end{figure}

We measure the radial evolution of the thickness, $h$,  at different times, $t$. Both actual values and normalized data, $H=h/d_0$ and $R=r/d_0$, are shown in figure~\ref{fig:fig4}. At short times, the liquid sheet is rather thick and its thickness decays very rapidly with the radial distance. The radial evolution becomes smoother as time evolves. At the maximum expansion ($t=4.57$ ms, $T=0.42$), the thickness is almost uniform: it increases from $35 \, \mu$m to $39 \, \mu$m for short distances and then decreases down to $26 \, \mu$m. In the retraction regime, the thickness is measured to vary non monotonically with the radial distance, but is always small. We will not consider the retraction regime in the following due to the absence of theoretical models and to the very restricted range of variation of the thickness in this regime. We will therefore only consider in the following data taken for $t<4.57$ ms ($T<0.42$). In the inset of figure~\ref{fig:fig4}, representative values plotted in a log/log scale indicate two scaling regimes at short and long times and a crossover between these two regimes at intermediate time.

\begin{figure}
\centerline{\includegraphics[height=16cm,width=13cm]{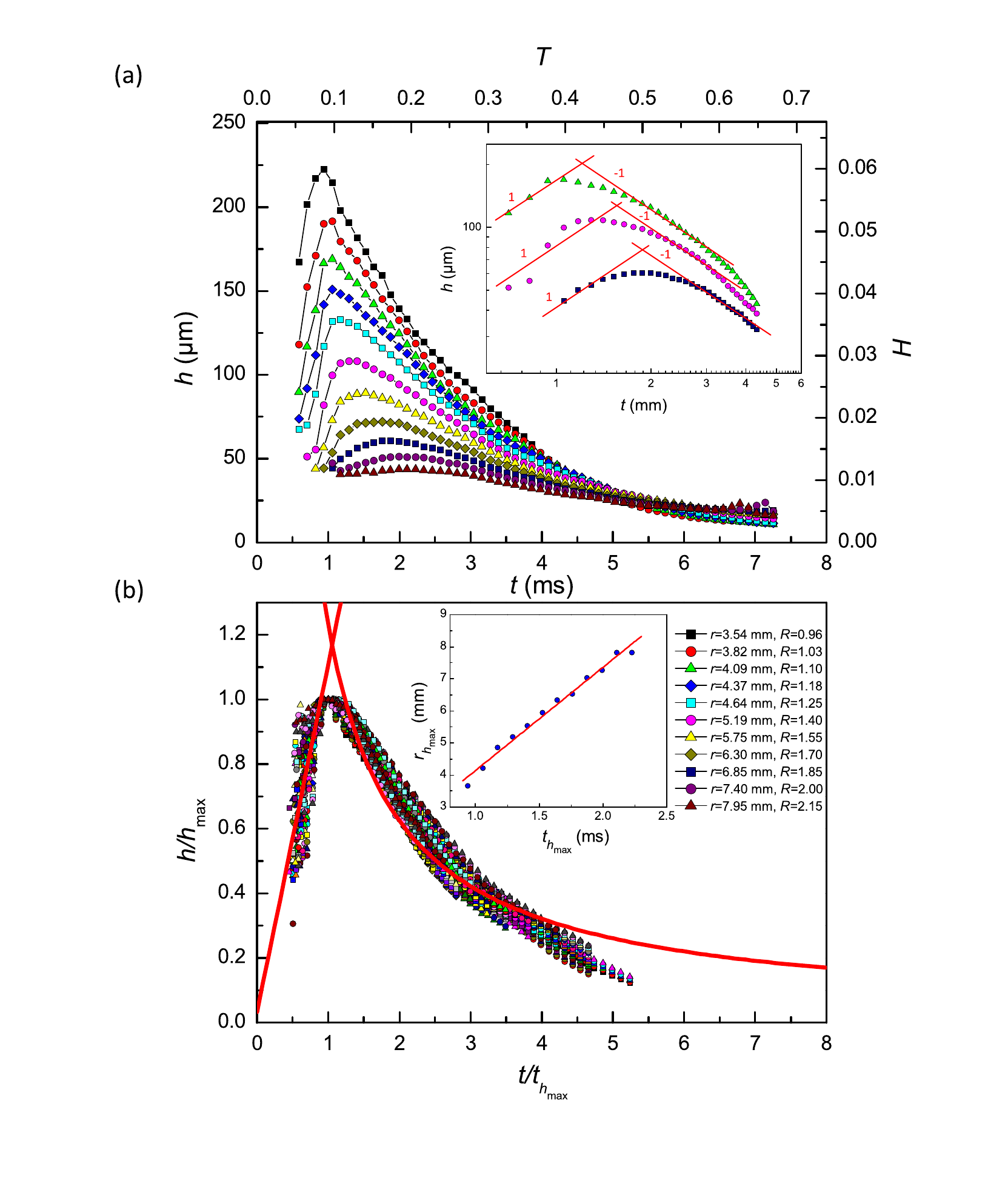}}
  \caption{(a) Time evolution of the thickness of a liquid sheet. Actual and normalized data are plotted for different radial positions, as indicated in the legend of (b). (Inset) Representative data in a log/log scale to highlight the two asymptotic scaling regimes. (b) Same data as those shown in (a), but normalized in the $x$ axis by the time at which the thickness is maximum, $t_{\rm{h_{max}}}$, and in the $y$ axis by the maximum thickness, $h_{\rm{max}}$. The lines correspond to the two asymptotic scaling regimes (as deduced from figure~\ref{fig:fig6}). Only data corresponding to the expansion regime are shown in (b). The drop diameter is $d_0=3.7$ mm and the impact velocity is $u_0=4.0$ m/s. (Inset) Variation of the radial position of the maximum thickness of the sheet as a function of time. The symbols are the experimental data and the line is the best linear fit of the data, yielding a velocity $r_{\rm{h_{max}}} / t_{\rm{h_{max}}} = 3.2$ m/s. }
\label{fig:fig5}
\end{figure}

It is more instructive to plot the time- and space-resolved data of the thickness as a function of time for different radial distances. This is shown in figure~\ref{fig:fig5}(a) for both natural and normalized data. For all radial positions, the thickness of the sheet exhibits a non monotonic evolution with time. We find that the maximum thickness occurs at longer times for larger radial positions, recalling the propagation of a shock-wave. Representative data in a log/log scale are shown in the inset of figure~\ref{fig:fig5}(a) and indicate two asymptotic regimes. Inset of figure~\ref{fig:fig5}(b) shows $r_{\rm{h_{max}}}$ versus $t_{\rm{h_{max}}}$, where $t_{\rm{h_{max}}}$ is the time at which the thickness is maximum at the radial position $r_{\rm{h_{max}}}$. We find a linear variation, demonstrating a constant velocity. Best fit of the data yields a speed $v_{\rm{h_{max}}} = 3.2$ m/s for the displacement of the maximum thickness. Interestingly, all the data acquired at different radial positions collapse on a single master curve once the thickness and the time are normalized by their values at the maximum thickness (fig.~\ref{fig:fig5}(b)), which highlights a universal shape for the evolution of the thickness with time and radial distance. In the following, we evaluate how the thickness scales with time and radial distances, both for the short-time regime where the thickness increases with time, and for the long-time regime where $h$ decreases with $t$.

\subsection{Scaling and comparison with theoretical predictions}

\begin{figure}
\centerline{\includegraphics[height=9.5cm,width=16cm]{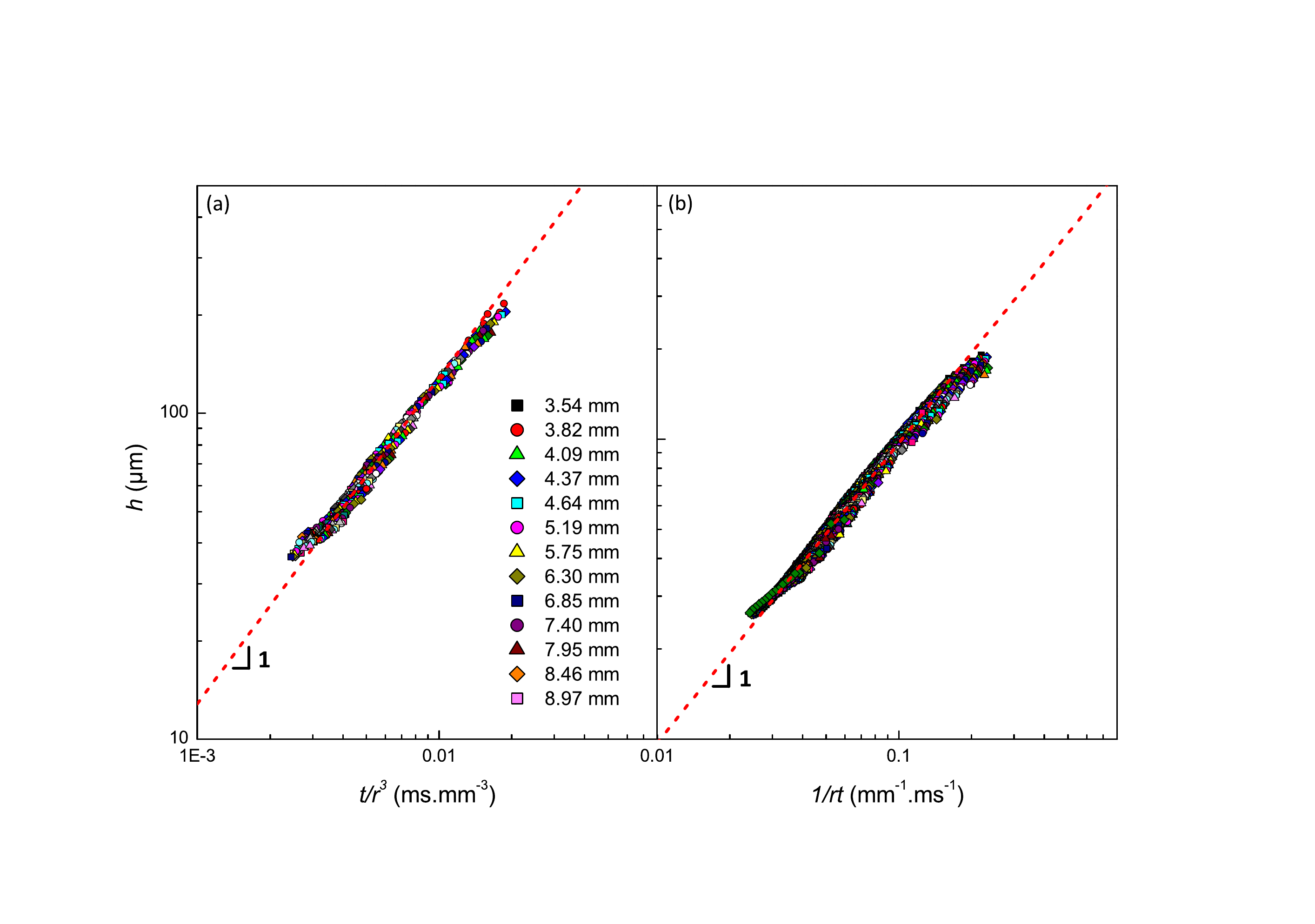}}
  \caption{Same data as in  figure~\ref{fig:fig5} but plotted as a function of $t/r^3$ (a) and $1/rt$ (b) where $r$ is the radial distance and $t$ is the time elapsed from the impact of the drop. The data corresponding to $t<t_{\rm{h_{max}}}$ are shown in (a), and those corresponding to $t>t_{\rm{h_{max}}}$ are shown in (b). Same symbols as in figure~\ref{fig:fig5}. The dashed lines are power law fits of the experimental data with a slope of $1$.}
\label{fig:fig6}
\end{figure}

\begin{figure}
\centerline{\includegraphics[height=9cm,width=16cm]{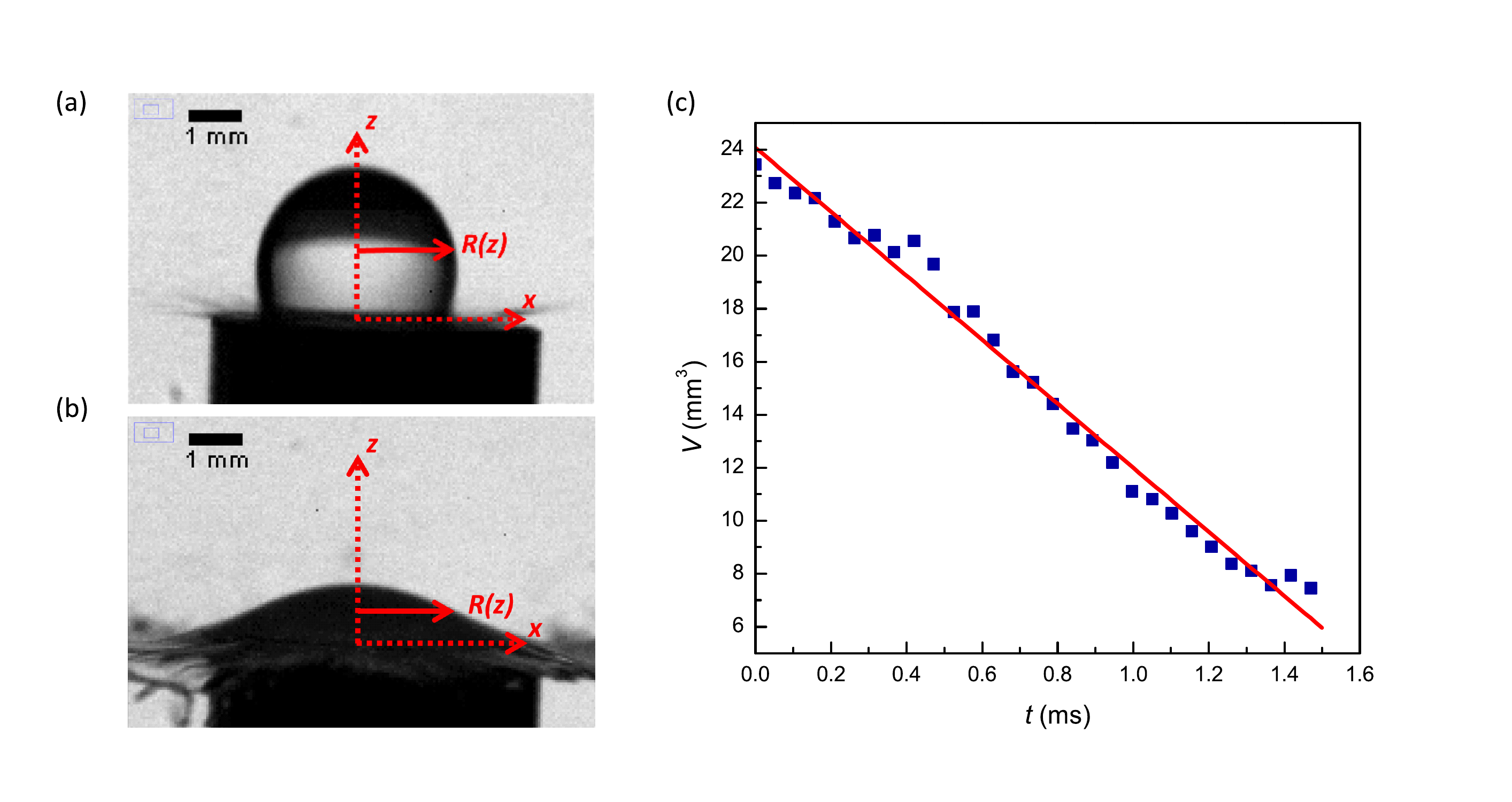}}
  \caption{(a,b) Side views of a liquid drop impacting a target, at time $t=0.10$ ms (a) and  $t=0.79$ ms (b). (c) Time evolution of the volume of liquid on the target. Symbols are experimental data and the continuous line is a linear fit yielding a constant liquid ejection rate. Here, the drop diameter is $d_0=3.7$ mm and the impact velocity is $u_0=4.0$ m/s.}
\label{fig:fig7}
\end{figure}

We first consider the short time regime where $t<t_{\rm{h_{max}}}$. In this regime, the thickness varies over almost one order of magnitude, ranging from $36 \, \mu$m to $ 217 \, \mu$m. We find that all data acquired at different radial positions and different times can be collapsed onto a single mastercurve if the thickness is plotted as a function of $t/r^3$ (fig.~\ref{fig:fig6}(a)). Moreover, a fit of the master curve (dashed line) shows that $h \propto t/r^3$ in this regime  as predicted by \citet{Rozhkov2004}. More specifically, \citet{Rozhkov2004} predicts

\begin{equation}
h = \alpha \, d_0^3 \, u_0 \, \frac{t}{r^3}
\label{eq:Hr}
\end{equation}

with $u_0$ the impact velocity of the drop and $d_0$ the diameter of the drop. A fit of the experimental data yields  $\alpha = 0.061$. This value is twice as large the value ($\alpha \approx 0.038$) derived by \citet{Rozhkov2004} for roughly equivalent experimental conditions. Note that in \citet{Rozhkov2004} $\alpha$ could be evaluated based on the observation of Mach-Taylor rupture waves generated by cutting radially the liquid sheet with a very thin obstacle, but not from thickness measurements. Hence, at short times, experimental data are well accounted by the model developed in \citet{Rozhkov2004}. This model is predicted to be valid at short times, up to $3$ times the typical collision time $\tau_{\rm{coll}}= d_{0}/ u_{0}= 0.93$ ms. Our data fall safely within this limit as, independently of the radial position, the maximum thickness occurs at times no longer than $2$ ms (inset fig.~\ref{fig:fig5} (a)). One key hypothesis of this model is that a point source, i.e. the impacting drop, feeds the sheet at a constant volume rate. We directly check this assumption thanks to side-view images of the drop impacting the target and feeding the liquid sheet (fig.~\ref{fig:fig7} (a,b)). We evaluate the time-evolution of the volume of the liquid drop still available on the target, assuming that the drop remains axisymmetric at all times. The volume $V$ of the liquid is measured to decrease linearly with time yielding a constant volume rate $Q = - \rm{d}\textit{V}/\rm{d}\textit{t} = 12.1$ mL/s, in relatively good agreement with the value ($Q \simeq 6$ mL/s) found indirectly in \citet{Rozhkov2004}.

On the other hand, in the late stage of the expansion regime, $t>t_{\rm{h_{max}}}$, the thickness varies between $26 \, \mu$m and $ 191 \, \mu$m. We find that all data acquired at different radial positions can be collapsed onto a single master curve if the thickness, is plotted as a function of $1/rt$ (figure~\ref{fig:fig6}(b)). We find that $h  \propto 1/rt$, as predicted by \citet{Villermaux2011}. More specifically, \citet{Villermaux2011} predicts

\begin{equation}
h = \beta  \frac{\pi d_0^3}{6 u_0} \frac{1}{rt}
\label{eq:Hv}
\end{equation}

A fit of the experimental data provides an estimate of the prefactor $\beta=0.142$. Note that this value is in very good agreement with the predicted value ($\beta=\frac{1}{2\pi}=0.159$).

Combining our findings for the scaling with time and position of the sheet thickness for the two stages of the expansion regime, one can provide a prediction for the maximum thickness position. By equalling the two functional forms found for the thickness, equation~\ref{eq:Hr} and  equation~\ref{eq:Hv}, one finds

\begin{equation}
v_{\rm{h_{max}}} = r_{\rm{h_{max}}}/t_{\rm{h_{max}}}=  \sqrt{\frac{6 \alpha}{\pi \beta}} \, u_0
\label{eq:Speed}
\end{equation}

First, equation~\ref{eq:Speed} shows that the position of the maximum thickness moves at a constant velocity with time, as found experimentally. Moreover, it predicts that the velocity is controlled by the impact velocity $u_0$. The proportionality constant between $v_{\rm{h_{max}}}$ and $u_0$ is $\sqrt{\frac{6 \alpha}{\pi \beta}}=0.906$. Hence, one predicts $v_{\rm{h_{max}}}= 3.6$ m/s (for $u_0=4.0$ m/s), in good quantitative agreement with the direct measurements ($v_{\rm{h_{max}}} = 3.2$ m/s). The good agreement can also be inferred from the normalized experimental data shown in figure~\ref{fig:fig5}(b), where the lines corresponding to fits of the data in the two regimes are found to cross very close to the point corresponding to $h=h_{\rm{max}}$ and $t=t_{\rm{h_{max}}}$.

\subsection{Results at different We numbers}

\begin{figure}
\centerline{\includegraphics[height=9cm,width=13cm]{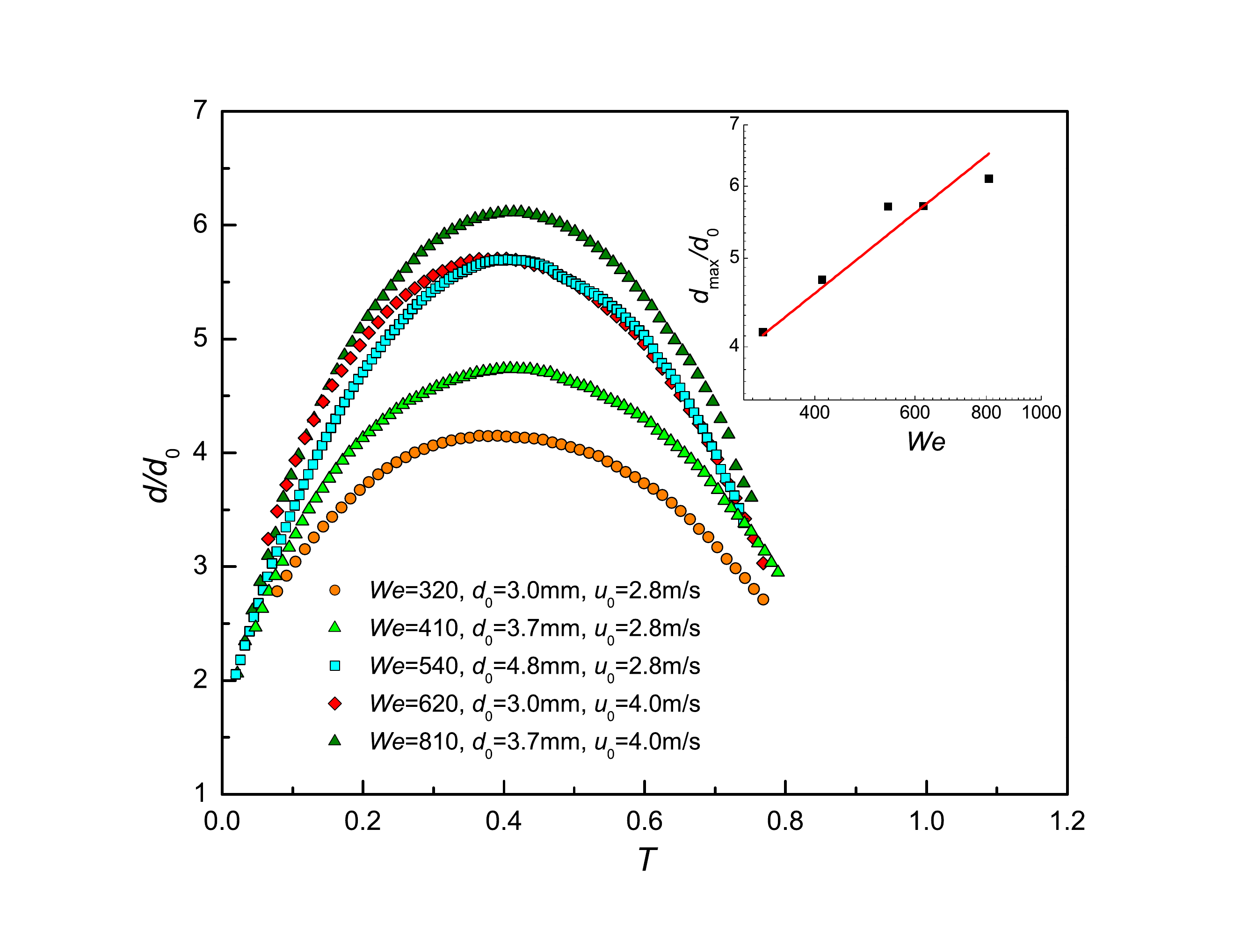}}
  \caption{Sheet diameter evolution for several experiments at different $We$ numbers as indicated in the legend. (Inset) Maximal expansion plotted as a function of $We$. Symbols are experimental data and the line is a power law fit with an exponent of $0.5$: $d_{\rm{max}}/d_0= \sqrt{We/19}$.}
\label{fig:fig8}
\end{figure}

To check the generality of our measurements, we perform the same type of measurements and analysis as detailed above, for different Weber numbers where the impact drop velocity and the volume of the drop are varied. Five different experimental conditions, for which the Weber numbers varies between $320$ and $810$, are performed. In all experiments, we vary the diameter of the target to keep constant ($0.6$) the ratio between the diameter of the drop and that of the target. Figure~\ref{fig:fig8} shows the time evolution of the diameter of the liquid sheet for the different experiments. The maximum extension of the sheet occurs at a time that depends on the drop diameter and not on the impact velocity. We find that the maximum extension occurs roughly at the same normalized time $T=0.40 \pm 0.01$. Here the standard deviation results from the statistics over the five experimental configurations. We measure that the sheet expands more as $We$ increases, from $d_{\rm{max}}/d_0=4.1$ to $d_{\rm{max}}/d_0=6.1$. Although on a very restricted range, we find that our data are compatible with a maximum expansion scaling with $(We)^{0.5}$, as observed and predicted theoretically for liquid sheets expanding in air \citep{Rozhkov2004, Villermaux2011} but also on solid surfaces  \citep{Eggers2010, Bennett1993, Marengo2011}. A fit of the data using the functional form $d_{\rm{max}}/d_0= \sqrt{We/K}$ gives $K=19$, in excellent agreement with the values ($K=20$) given in \cite{Rozhkov2004} for similar experimental conditions.

\begin{figure}
\centerline{\includegraphics[height=14cm,width=17cm]{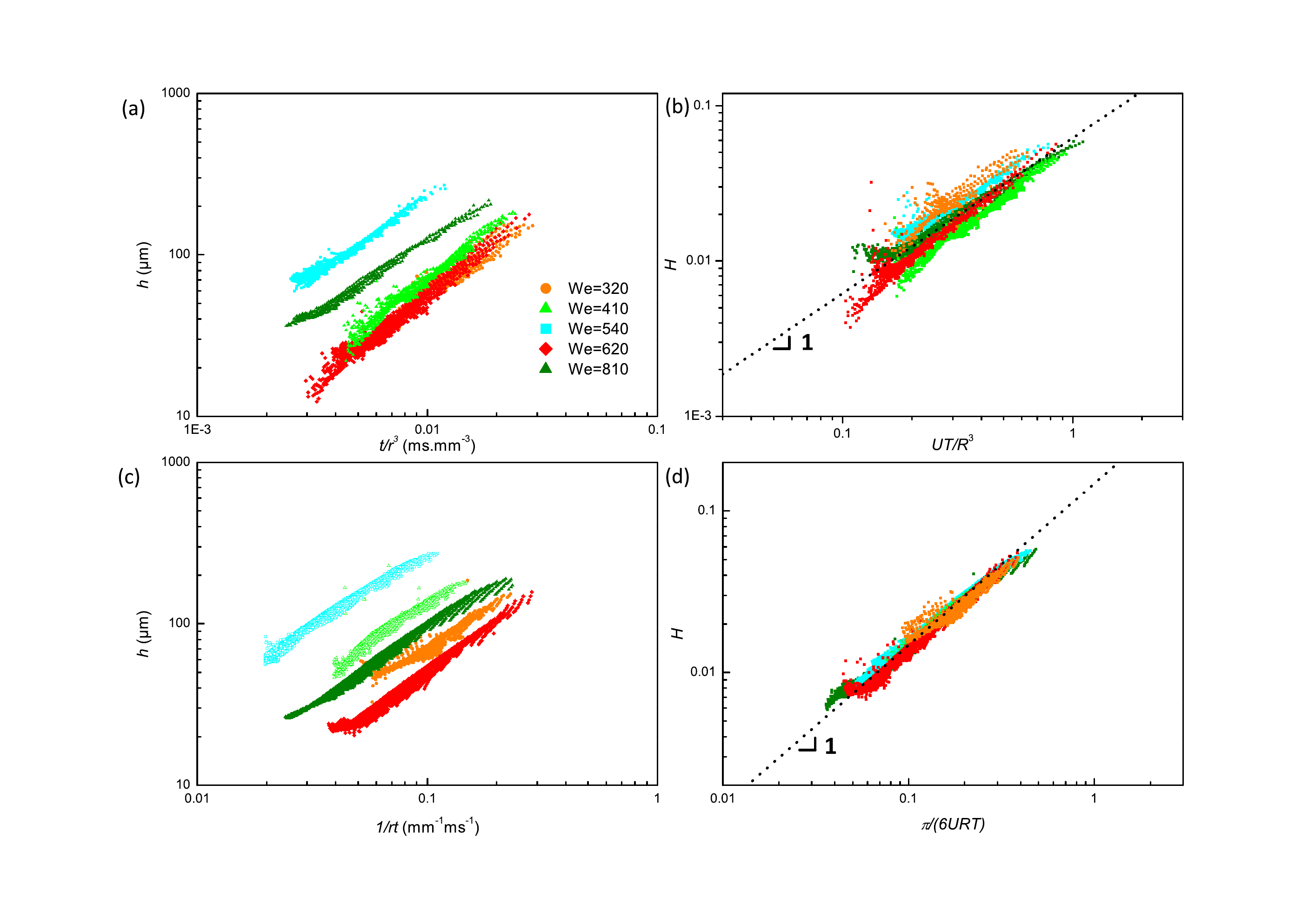}}
  \caption{Evolution with time, $t$, and radial distance, $r$, of the thickness of a liquid sheet for five different experimental configurations as indicated in the legend. (a, b) correspond to data in the early stage expansion regime ($t<t_{\rm{h_{max}}}$), and (c,d) correspond to data in the late stage expansion regime ($t>t_{\rm{h_{max}}}$). (a, c) Actual thickness plotted as a function of (a) $t/r^3$ and (c) $1/rt$. (b, d) Normalized thickness, $H$, plotted as a function of (b) $UT/R^3$,  and (d) $1/URT$. In (b, d), the dashed lines are power law fits of the experimental data with a slope of $1$.}
\label{fig:fig9}
\end{figure}

The raw data of the sheet thickness in the short time regime and long time regime of the expansion stage are shown in figure~\ref{fig:fig9}(a) as a function of $t/r^3$ and in figure~\ref{fig:fig9}(c) as a function of $1/rt$. Overall the thickness varies between $12 \, \mu$m and $272 \, \mu$m, and the maximal thicknesses are measured for the experiments performed with $d_0=4.8$ mm and $u_0=2.8$ m/s ($We=538$). In all cases a good scaling of the data acquired at different radial positions is obtained, showing the robustness of our findings. Using as normalized data, $H=h/d_0$, $R=r/d_0$, $T=t/\tau$ and $U=\frac{u_0 \tau}{d_0}$, equation~\ref{eq:Hr} for the short-time expansion regime ($t<t_{\rm{h_{max}}}$) rewrites

\begin{equation}
H = \alpha \, \frac{UT}{R^3}
\label{eq:HrNorm}
\end{equation}

and equation~\ref{eq:Hv} for the late time expansion regime ($t>t_{\rm{h_{max}}}$) rewrites

\begin{equation}
H = \beta  \frac{\pi}{6} \frac{1}{URT}
\label{eq:HvNorm}
\end{equation}

Remarkably, figures~\ref{fig:fig9}(b) and (d) show a nice collapse of the data acquired  for different Weber numbers once plotted in a normalized fashion. A fit with a powerlaw with an exponent $1$ of the data in the early expansion regime (fig.~\ref{fig:fig9}(b)) gives $\alpha= 0.062$, and a fit with a powerlaw with an exponent $1$ of the data in the late stage of the expansion regime (fig.~\ref{fig:fig9}(d)) gives $\beta= 0.144$. By equating equations~\ref{eq:HrNorm} and  ~\ref{eq:HvNorm}, one predicts a constant normalized velocity for the maximum thickness $V_{\rm{h_{max}}}=(R/T)_{\rm{h_{max}}}=\sqrt{\frac{6 \alpha}{\pi \beta}} U = 0.91 U$. By directly measuring the normalized velocity of the thickness maximum for the different experiments, we find $V_{\rm{h_{max}}}= (0.70 \pm 0.18) U$, in good agreement with the previous determination.

\subsection{Azimuthal thickness modulations}\label{Modulation}

\begin{figure}
\centerline{\includegraphics[height=10cm,width=15cm]{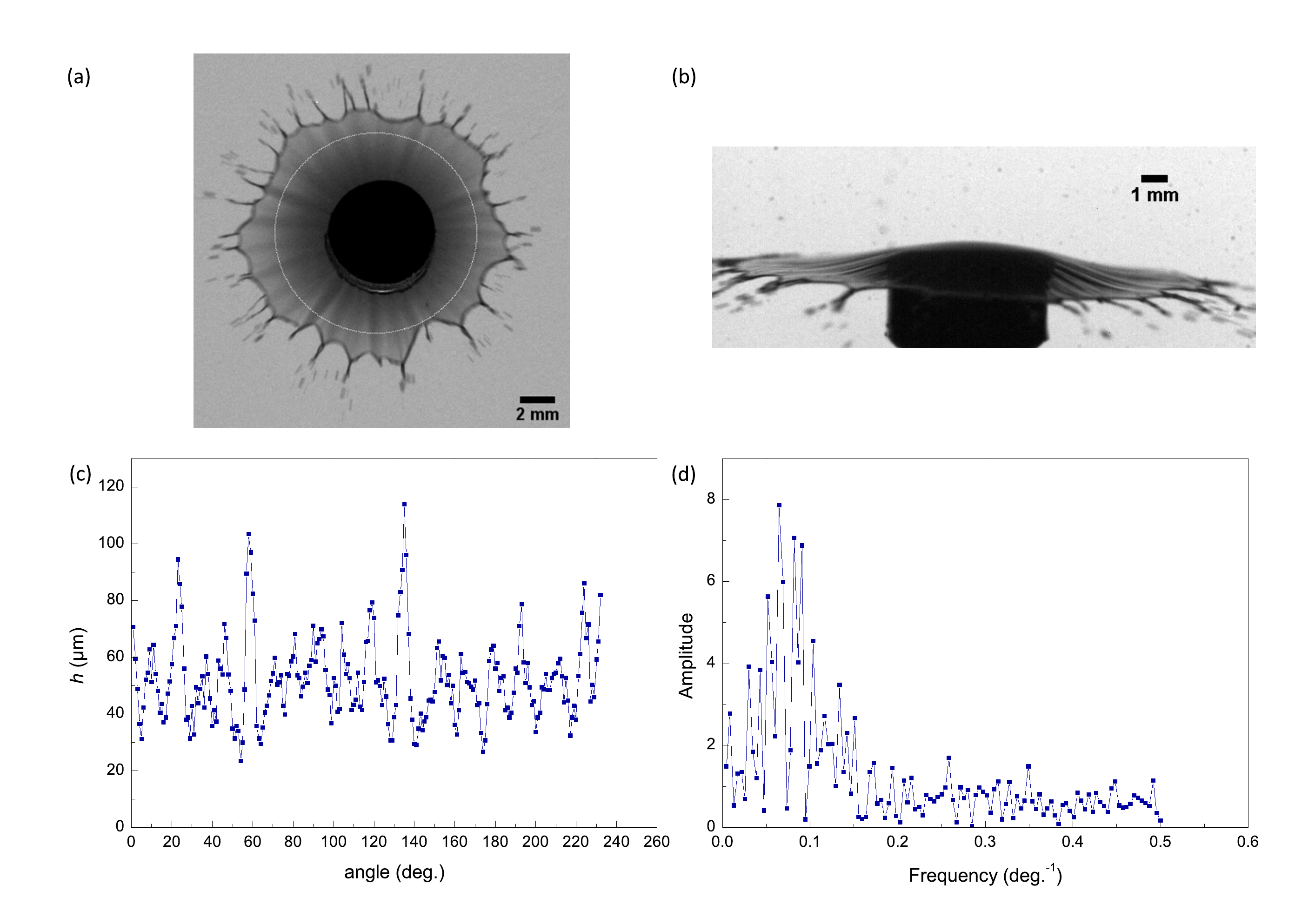}}
  \caption{Front (a) and side (b) views of liquid sheets showing a modulation of the thickness. In (a) (resp. (b)) $t=1.03$ ms, $T=0.064$  (resp. $t=1.44$ ms, $T=0.09$). (c) Azimuthal profile of the thickness along the circle shown in (a). (d)   Fourier transform of the profile shown in (c). Here, the drop diameter is $d_0=3.7$ mm and the impact velocity is $u_0=4.0$ m/s}
\label{fig:fig10}
\end{figure}

Thanks to the presence of a dye in the sheet, we are able to directly visualize small fluctuations of the thickness of the liquid sheet. We have previously integrated the values of the thickness measured at different azimuthal angles to get averaged data that just depend on the radial distance. A careful examination of the liquid sheet shows however a clear azimuthal modulation of the thickness. Thicker than the average, channels are observed to run over the whole extension of the sheet (fig.~\ref{fig:fig10}(a)). The profile of the thickness along a circle centered at the center of the target clearly shows fluctuations of the thickness of typical amplitude $15 \, \mu$m (hence about $30 \%$ of the average thickness ($52 \,\mu$m)). Moreover the channels are rather regularly arranged as evidenced by a peak in the Fourier transform of the profile (fig.~\ref{fig:fig10}(d)), indicating a characteristic spacing of $15$ deg. between adjacent channels. We are not aware of any experimental or theoretical studies of such thickness modulations. We were not able however to see any clear correlation between the thickness modulation and the rim corrugation. Note that the thickness modulation can also be inferred from the side view of a sheet, as shown in figure~\ref{fig:fig10}(b). We believe that these phenomena would deserve further investigations.

\section{Conclusion and perspectives}
\label{Conclusion and perspectives}

We have described a quantitative yet simple method to measure the thickness field $h(r,t)$ of a free liquid sheet developing in air  with a temporal resolution set by the recording camera and a spatial resolution set by the pixel size. Measurable thicknesses lie in the range $10-450 \mu \textrm{m}$. This method has been applied to the analysis of the  radial expansion of a free liquid sheet resulting from the impact of a  falling drop on a small target and has allowed one to evidence new features:

(i) Two asymptotic hydrodynamic regimes characterize the expansion of the liquid sheet. In the first regime observed at short time,  $h(r,t)\propto t/r^3$ corresponding to the feeding of the sheet by a central point source reminiscent of the  impacting drop in agreement with \cite{Rozhkov2004}. The later regime of expansion corresponds to a pseudo-stationary regime close to the hydrodynamic regime of a Savart's sheet where $h(r,t)\propto 1/(rt)$ as predicted by \cite{Villermaux2011}.

(ii) Experimentally, we observe for each radial position a maximum of the thickness $h_{\rm{max}}$ with time, which propagates at a constant celerity,  close to the impact velocity of the drop. This maximum  as well as  its constant propagation celerity can be theoretically inferred from the cross-over between the two asymptotic regimes.

(iii)  An azimuthal modulation of the thickness is observed during the expansion of the sheet, but it seems not directly at the origin of the radially expelled ligaments, which are localized at the rim and result from a Rayleigh-Taylor-like instability.

Our robust and simple experimental method for thickness measurement  could be  advantageously used  to investigate other physical situations where free liquid  sheets develop, as for instance in water bells (\cite{Clanet2007}) or fan jets that exit from the nozzles used for agricultural sprays \citep{Bergeron2003}. In this latter case, it has been shown for a long time that sheets of complex fluids including dilute emulsions \citep{Dombrowski1954, Ellis1997} and surfactants solutions \citep{Rozhkov2006}, or dilute colloidal suspensions \citep{Addo-Yobo2011}, can be destabilized by a perforation mechanism that is drastically different from the standard destabilization mechanism of a liquid sheet. Here, the sheet is punctuated by holes that nucleate and grow until resulting in a web of ligaments. The perforation mechanism has been found to be correlated with a reduction of the volume of small drops in the spray and is presently used to formulate agricultural spay liquids that exhibit so-called spray drift reduction features (\cite{Hilz2013a}). However, the physical mechanisms at the origin of the perforation process  remains poorly understood and controversial. Note that similar perforation processes have also been observed with liquid sheets of gaseous water (\cite{Lhuissier2013}). The capability of measuring the thickness field of sheets of complex fluids before and during the perforation instability,  also in model experiments as those described in the present paper,  is potentially  very attractive to answer such open questions.
 \\

We thank Emmanuel Villermaux and Jean-Christophe Castaing  for fruitful discussions, Adrian-Marie Philippe for providing us with his Matlab software and Pascal Martinez for help in building the experimental set-up. Financial support from Solvay is acknowledged.

\bibliographystyle{jfm}

\clearpage

\bibliography{JFM_14_S_0603_final}

\end{document}